\documentclass[twocolumn,times]{aastex62}
\usepackage{amsmath}
\usepackage{amssymb}
\usepackage{lineno}

\usepackage{bm}
\allowdisplaybreaks[4]

\def\be {\begin{equation}}
\def\ee {\end{equation}}

\def\tt {\{t\}}

\usepackage{color}

\shorttitle{Origin of Earth-Moon isotopic similarity}
\shortauthors{Wenshuai Liu}
\begin{document}

%\title{Slow radial migration of a gap-opening planet triggered by dust feedback}
\title{\large{\textbf{Origin of the Moon's Earth-like isotopic composition from giant impact on a differential rotating proto-Earth}}}

\correspondingauthor{Wenshuai Liu}
\email{674602871@qq.com}

\author{Wenshuai Liu}
\affiliation{School of Physics, Henan Normal University, Xinxiang 453007, China}



\begin{abstract}
According to the giant impact theory, the Moon formed by accreting the circum-terrestrial debris disk produced by Theia
colliding with the proto-Earth. The giant impact theory can explain most of the properties of the Earth-Moon system, however, simulations of giant impact between a planetary embryo and the growing proto-Earth indicate that the materials in the circum-terrestrial debris disk produced by the impact originate mainly from the impactor. Thus, the giant impact theory has difficulty explaining the Moon's Earth-like isotopic compositions. More materials from the proto-Earth could be delivered to the circum-terrestrial debris disk when a slightly sub-Mars-sized body collides with a fast rotating planet of rigid rotation but the resulting angular momentum is too large compared with that of the current Earth-Moon system. Since planetesimals accreted by the proto-Earth hit the surface of the proto-Earth, enhancing the rotation rate of the surface of the proto-Earth. The surface's fast rotation rate relative to the slow rotation rate of the inner region of the proto-Earth leads to transfer of angular momentum from surface to inner, resulting in the differential rotation. Here, we show that the giant impact of a sub-Mars-sized body on a differential rotating proto-Earth with a fast rotating outer region and a relative slow rotating inner region could result in a circum-terrestrial debris disk with materials predominately from the proto-Earth without violating the angular momentum constraint. The theory proposed here may provide a viable way of explaining the similarity in the isotopic compositions of the Earth and Moon.
\end{abstract}

\keywords{Earth-moon system --- giant impact --- smoothed particle hydrodynamics}


\section{Introduction}

The giant impact theory \citep{1,2} for the lunar origin is thought to lead to the formation of the Moon through accreting material from a circum-terrestrial debris disk after the impact of a planet-sized object with the proto-Earth, and numerical simulations with respect to this theory have found that the compositions of the circum-terrestrial debris disk where material eventually aggregates to form the Moon are predominately from the material of the impactor \citep{3,4,5}. Due to the fact that different Solar System bodies have distinct compositions, the compositions of the Moon would be significantly different from that of Earth according to the canonical giant impact theory. However, analysis of isotopic compositions from Earth and Moon indicates that compositions from Earth and Moon are highly similar \citep{6,7,8,9,10,11}, which is referred to as the isotopic crisis. In order to explain Earth-Moon isotopic similarity, the Moon should equilibrate with or form from the Earth's mantle after the impact \citep{12,13}, or form from giant impact with an impactor with identical isotopes as Earth \citep{14}. However, these models could not satisfy all of the geochemical observations.

\cite{15} proposed a modified model of giant impact with a fast-spinning proto-Earth. In this model, more material can be delivered to the circum-terrestrial debris disk which is made of material predominantly from the proto-Earth and massive enough to form the Moon after the collision of a slightly sub-Mars-sized body with a fast-spinning proto-Earth of rigid rotation. This theory can potentially satisfy the observation of isotopic composition but results in too large angular momentun compared with that of the recent Earth-Moon system. Other high energy, high angular momentum models include Synestia \citep{16,17} and a collision between two similar-mass bodies \citep{18}.

Simulations of terrestrial planet formation show that the final rigid rotation period can be faster than 4 hours \citep{19,20,21} after accretion of planetesimals and multiple giant impacts. Due to the fact that planetesimal accreted by the proto-Earth or multiple giant impacts on proto-Earth hit the surface of the proto-Earth, thus, enhance the rotation rate of the surface of the proto-Earth. The surface's fast rotation rate relative to the slow rotation rate of the inner region of the proto-Earth gives rise to angular momentum transfer from surface to inner, resulting in the differential rotation of the proto-Earth. In this work, with a differential rotating proto-Earth with a fast rotating outer region and a relative slow rotating inner region, the collision of a sub-Mars-sized body with such differential rotating proto-Earth could produce a  circum-terrestrial debris disk with material predominately from the proto-Earth without violating the angular momentum constraint.

We describe the mechanism in Section 2 and the discussions are in Section 3.

\section{Giant impact on a differential rotating proto-Earth}
Giant impacts are usually investigated using smoothed particle hydrodynamics (SPH). In these simulations, the target planet and impactor planet are modeled with particles whose motions are dominated by gravity and material pressure. In this section, the initial conditions for the planets are described and the results are given.

The impact scenario shown in Figure 1 is specified by the impact parameter $b=\sin(\beta)$ and the speed at first contact of the impactor with the target's surface $v_c$. We set the initial position of the impactor in order to let the contact occur 1 hour after the simulation's start. $\beta$ is set to be $17.5^o$ and $v_c$ to be $2v_{esc}$ where $v_{esc}=\sqrt{2G(M_t+M_i)/(R_t+R_i)}$ is the mutual escape speed of the system. Here, we set $M_t=1.05M_E$ and $M_i=0.05M_E$ where $M_E$ is the mass of the Earth.

To find the radius of the non-rotating target and the non-rotating impactor, the target and impactor are differentiated into a rocky mantle and an iron core with $70\%$ and $30\%$, respectively. In real situation, the outer region of the proto-Earth may be in the form of magma ocean after accretion of planetesimals and multiple giant impacts. Here, for simplicity, ANEOS $\mathrm{Fe}_{85}\mathrm{Si}_{15}$ and forsterite equations of state \citep{22} are used to model the materials. After getting the density profile of the non-rotating case \citep{23}, the target profile with rotation period of 4 hours are generated using WoMa \citep{23}.

In order to describe the differential rotating target for simplicity, as shown in the left panel of Figure 2, we rudely use the target density profile with rotation period of 4 hours generated by WoMa while setting the rotation period of the particles placed to precisely match the resulting density profiles to be 4 hours if $R\le4000\mathrm{km}$ and the rotation period to be 2.8 hours if $R>4000\mathrm{km}$ where $R$ is the distance between the particle and the target's center. For comparison, the giant impact on a rigid rotation target with rotation period of 2.8 hours is also conducted.

\begin{figure}
            \includegraphics[width=0.5\textwidth]{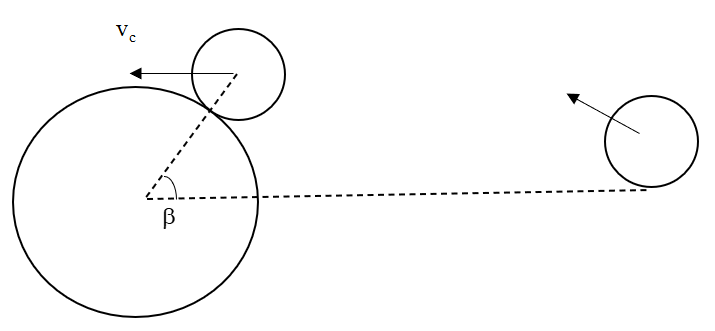}
\caption{The big circle represents the target planet with mass $M_t$ and the small circle is the impactor planet with mass $M_i$. $v_c$ is the speed at first contact of the impactor with the target's surface. The target planet rotates clockwise.}
\label{fig:figure1}
\end{figure}

\begin{figure}
     \begin{tabular}{cc}
            \includegraphics[width=0.17\textwidth]{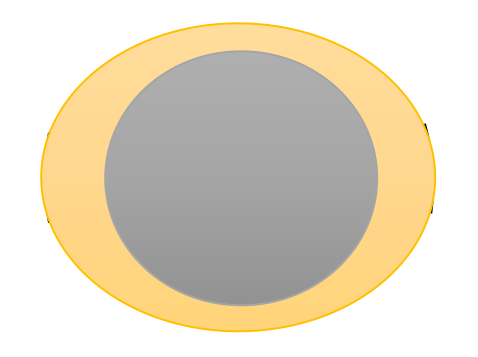}
            \includegraphics[width=0.17\textwidth]{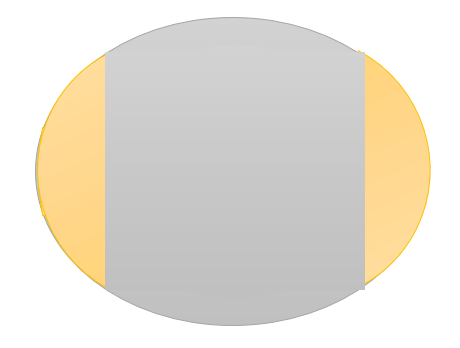}
            \includegraphics[width=0.17\textwidth]{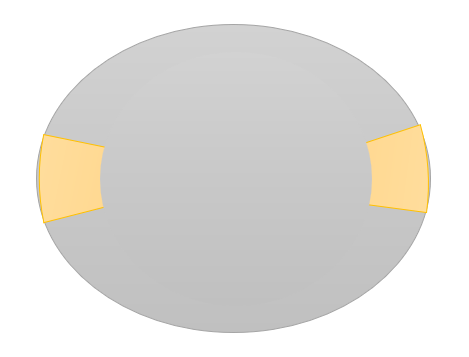}
            \end{tabular}
\caption{The differential rotating proto-Earth where the grey region and yellow region have different rotation period. The rotation axis is in the vertical direction}
\label{fig:figure2}
\end{figure}

\begin{figure*}
   \begin{center}
     \begin{tabular}{cc}
            \includegraphics[width=0.25\textwidth]{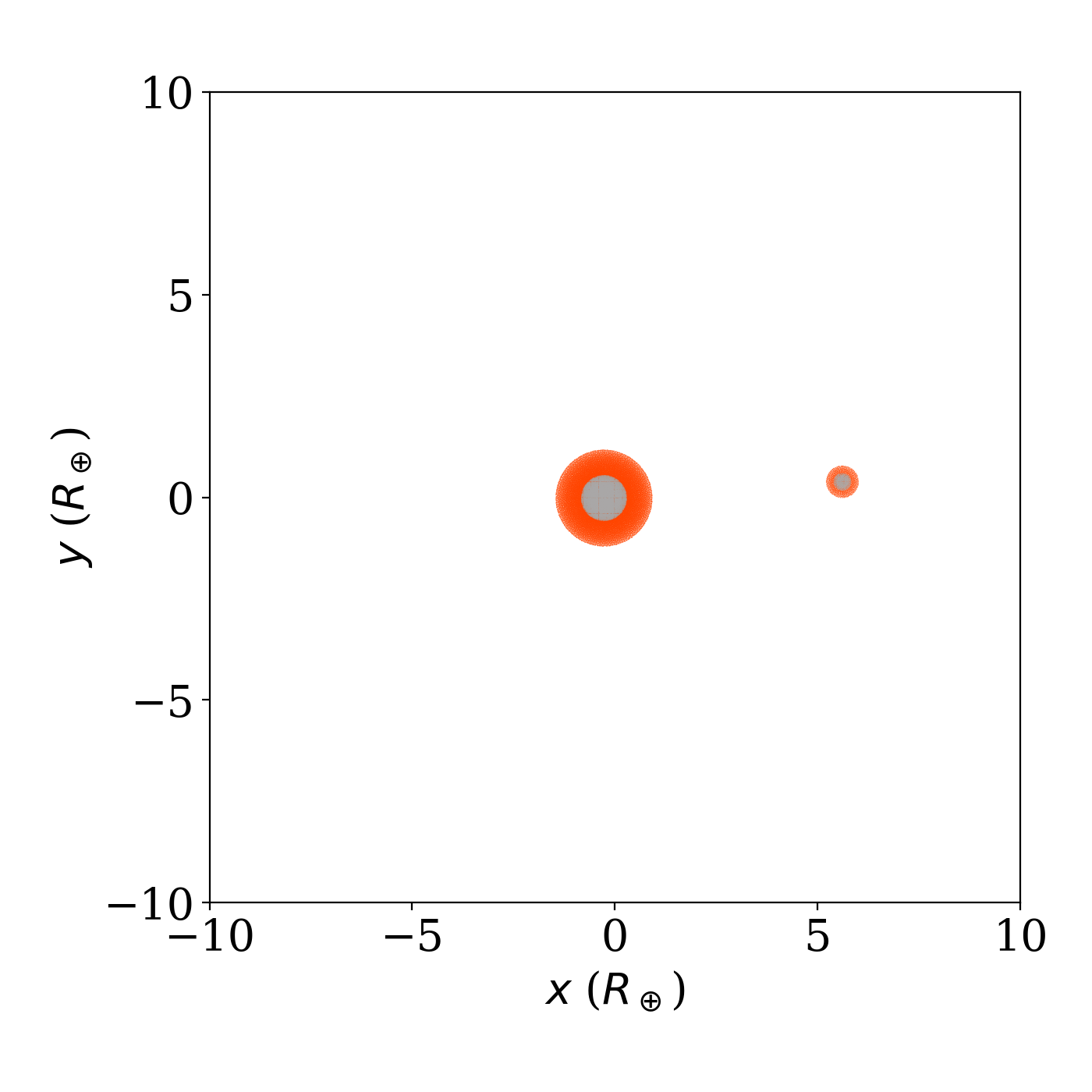}
            \includegraphics[width=0.25\textwidth]{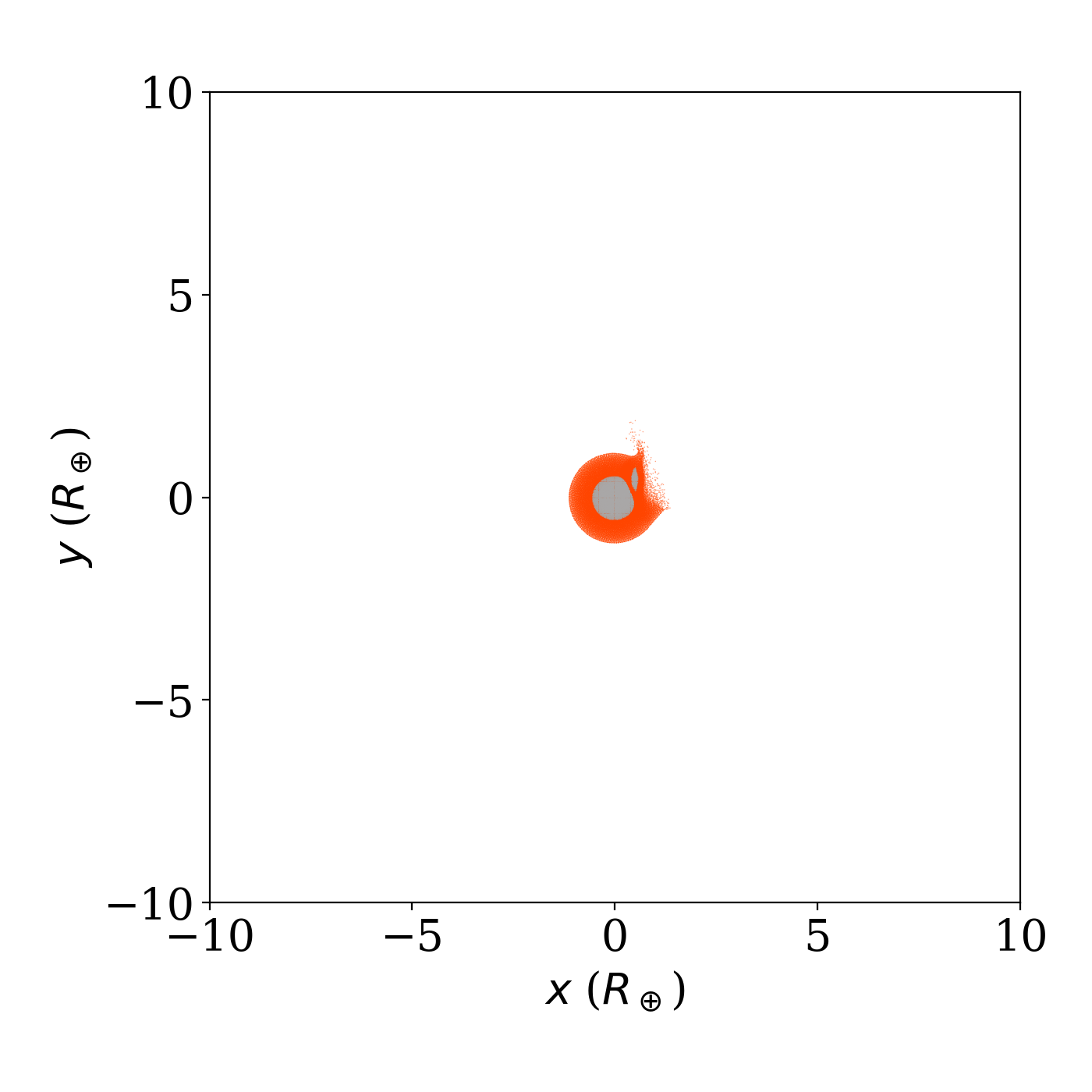}
            \includegraphics[width=0.25\textwidth]{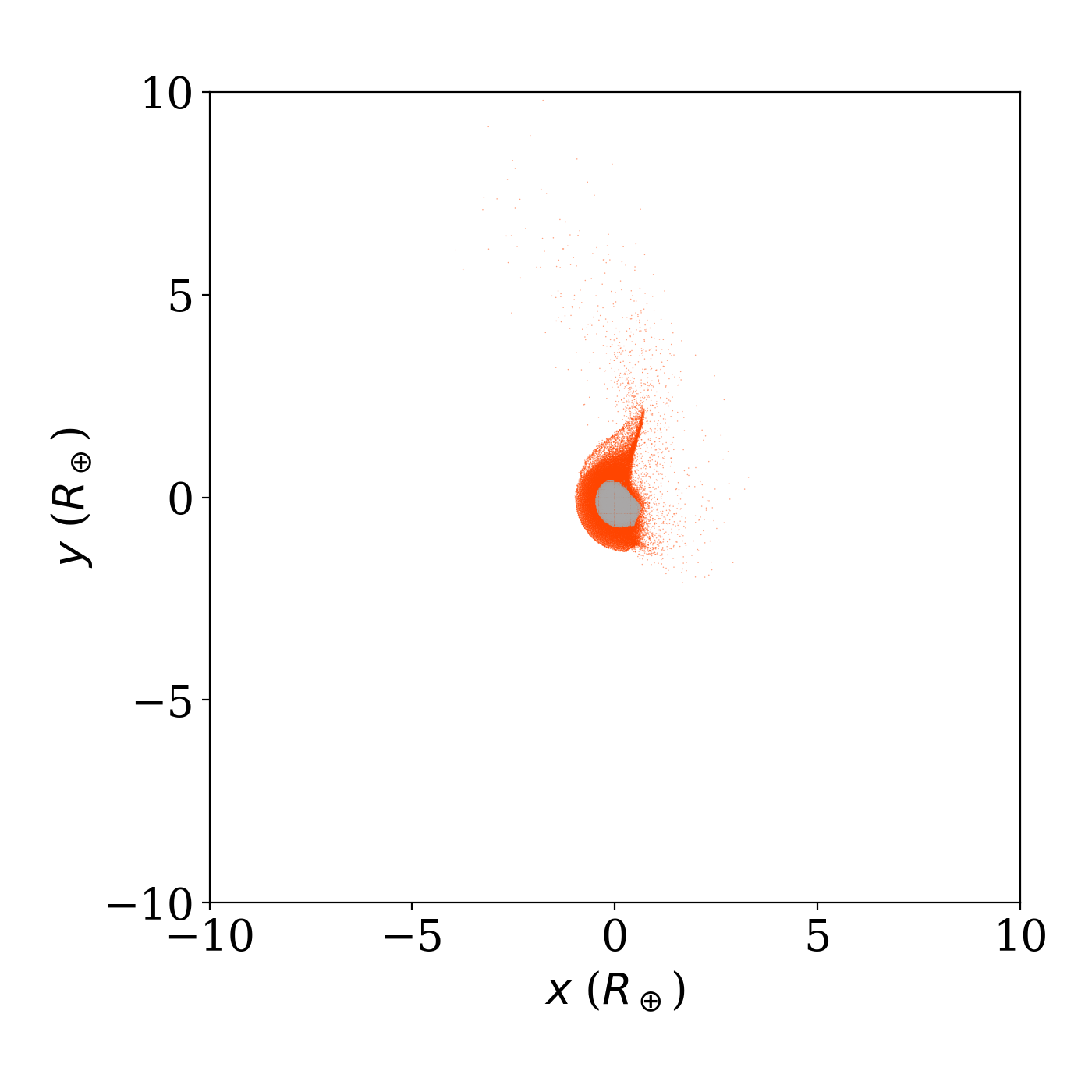}
            \includegraphics[width=0.25\textwidth]{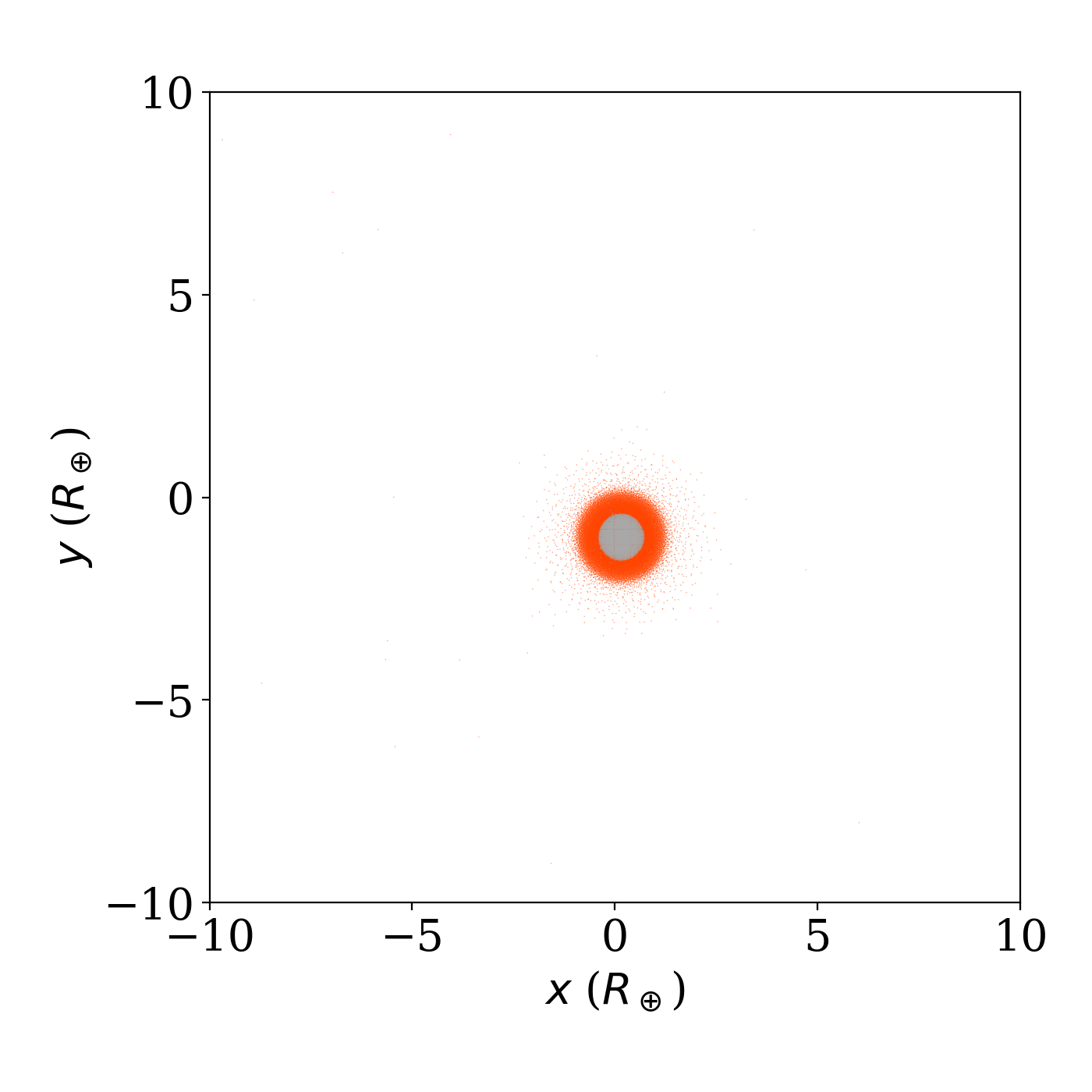}\\
            \includegraphics[width=0.25\textwidth]{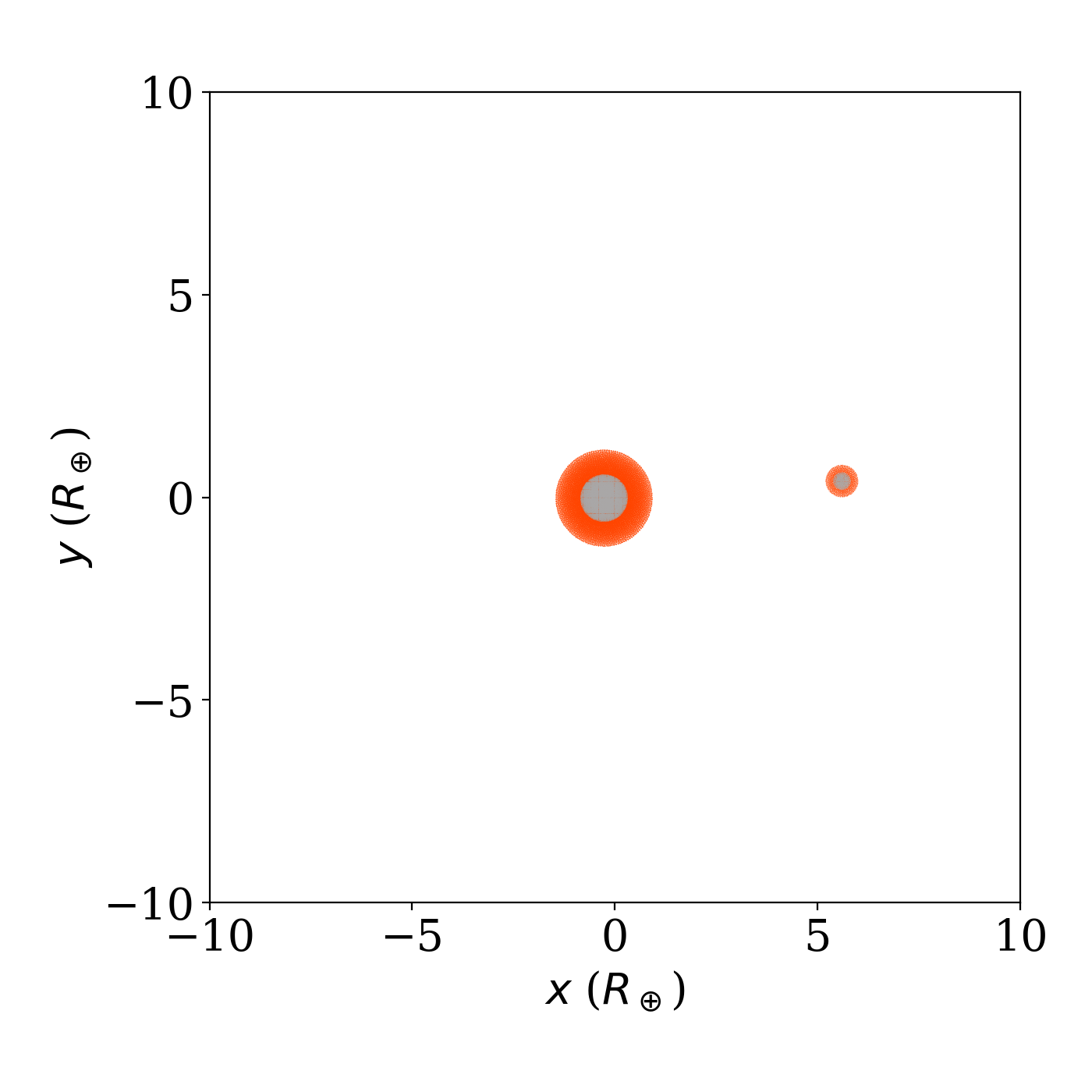}
            \includegraphics[width=0.25\textwidth]{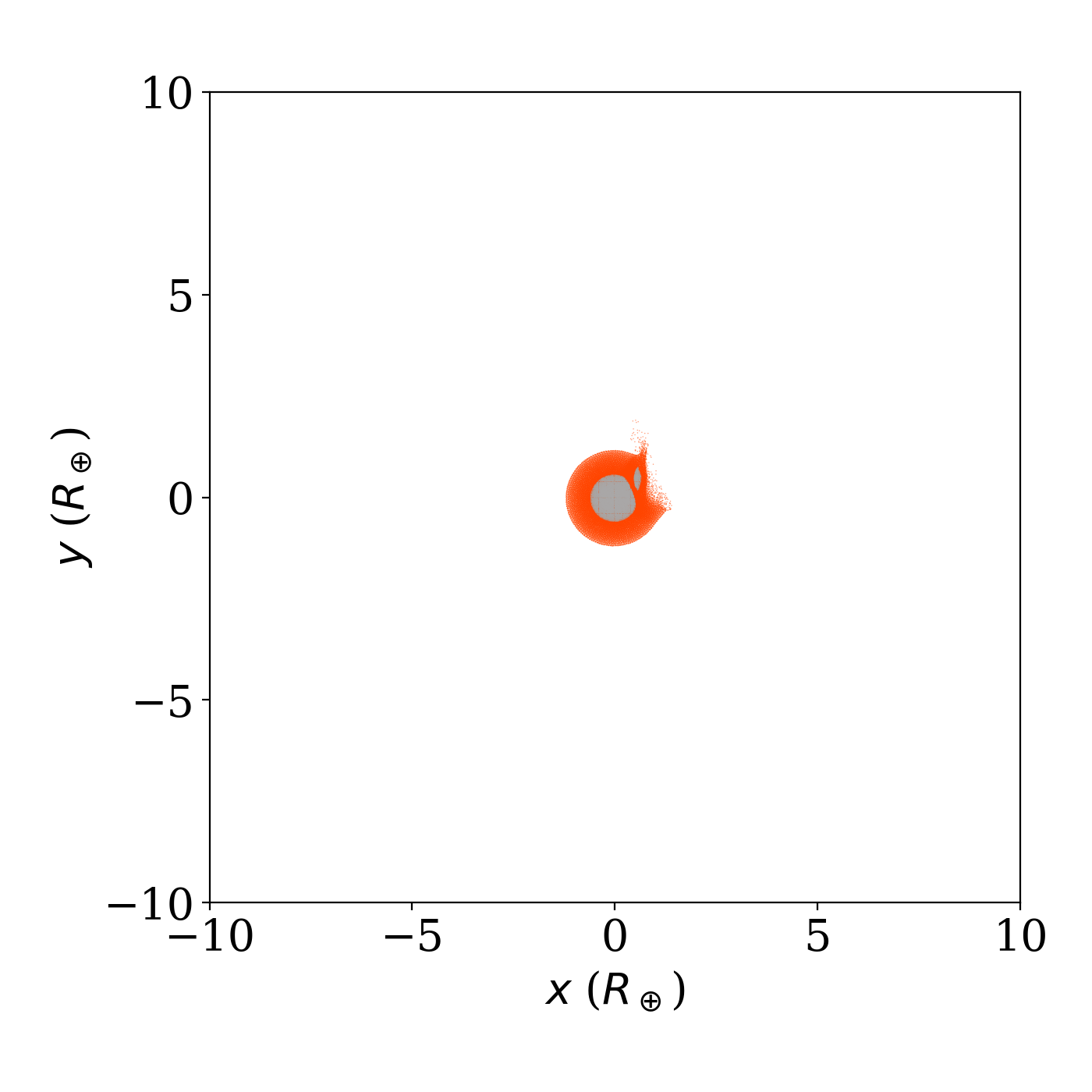}
            \includegraphics[width=0.25\textwidth]{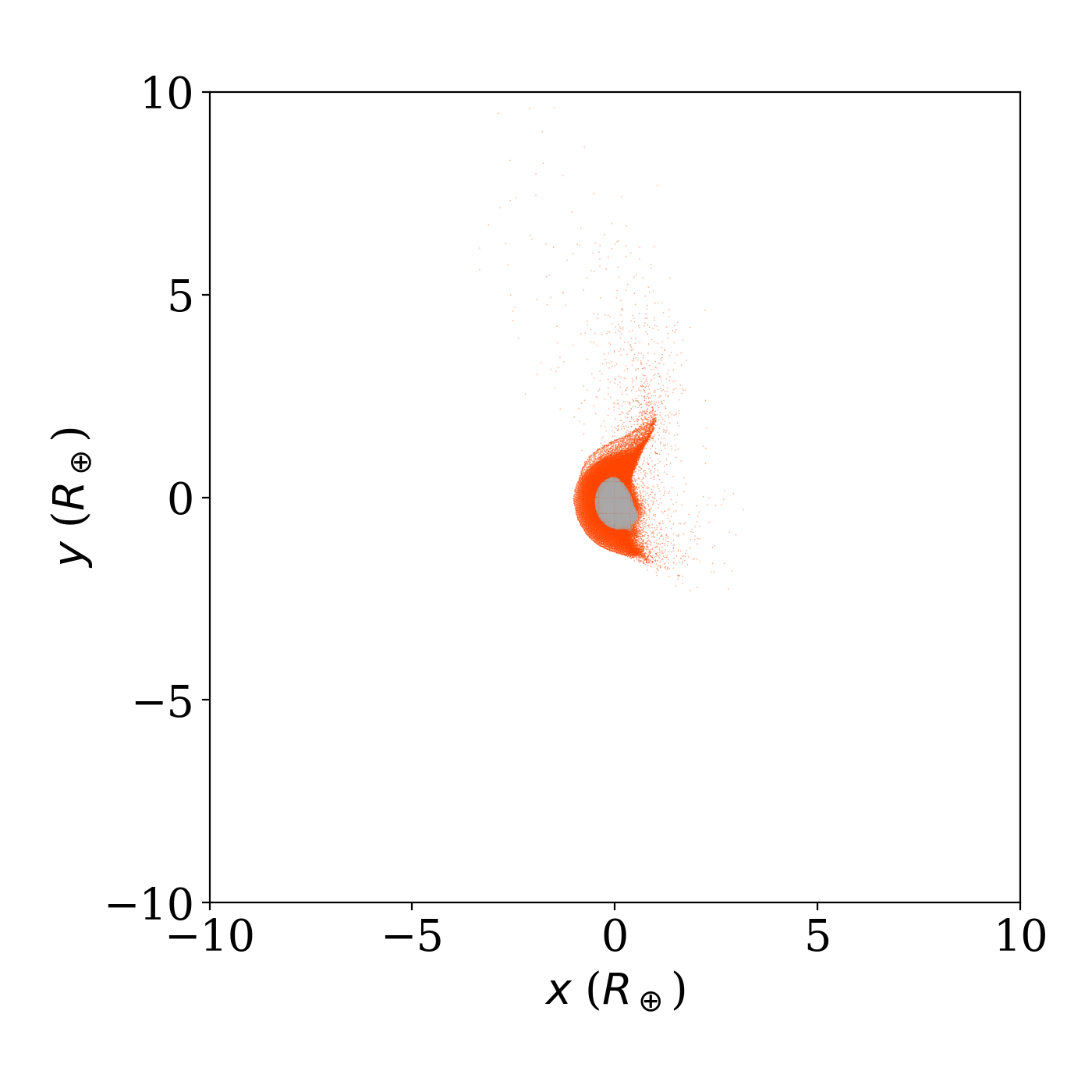}
            \includegraphics[width=0.25\textwidth]{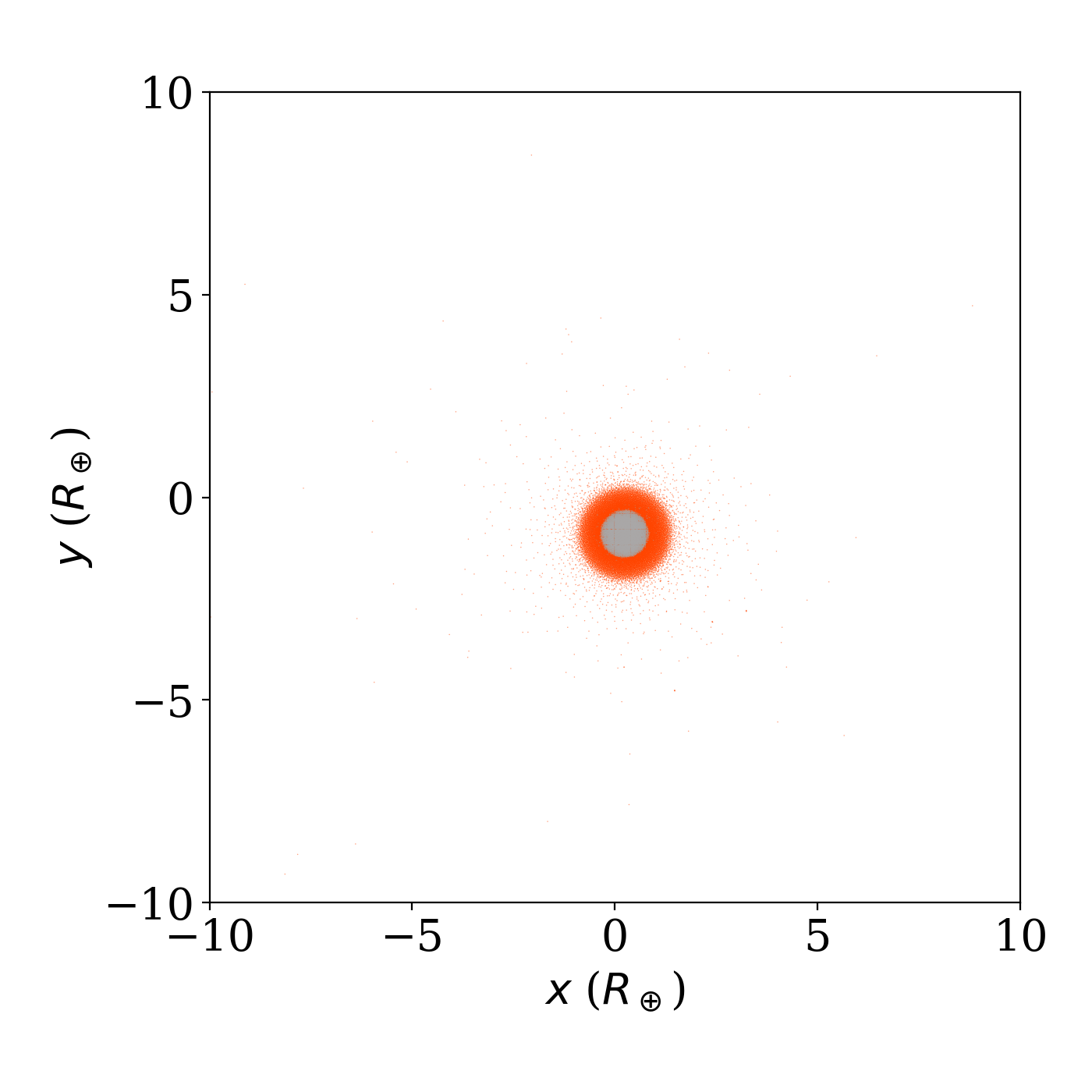}
            \end{tabular}
   \end{center}
\caption{Up and bottom panel represent the giant impact on a differential rotating planet and a rigid rotating planet, respectively. From left to right, the results are shown with simulation times of 0.56 hr, 1.11 hr, 1.67 hr and 30 hr.}
\label{fig:figure3}
\end{figure*}

With the condition above, we simulated the giant impact using code SWIFT \citep{24} with $10^6$ equal-mass SPH particles, the results are shown in Figure 3. It shows from Figure 3 that the results from these two simulations are similar. We further find that the mass of the debris disk generated from giant impact on the differential rotating proto-Earth is identical to that produced by giant impact on the proto-Earth with rigid rotation and that the materials in the debris disk from both simulations are mainly from the target. If the outer region is in the form of magma ocean, the results may be more obvious \citep{25}. Since the inner region of the differential rotating proto-Earth has a low angular velocity while the outer region has a large angular velocity, this means that, with suitable parameters of angular velocity in the inner and outer region with suitable radius, giant impact on a differential rotating proto-Earth with a faster rotating outer region may explain the Earth-Moon isotopic similarity while not violating the angular momentum constraint.

It should be noted that, due to angular momentum transfer from the outer region to the inner region of a terrestrial planet formed by accretion of planetesimals, the detailed profile of angular velocity is complex. In this work, we take the left panel of Figure 2 as the model of a differential rotating planet for simplicity. We expect the model of the middle and the right panel in Figure 2 may show similar results.

The theory proposed here is in some sense similar to that in \cite{15} and, unlike a rigid rotation adopted in \cite{15}, the proto-Earth just has a fast rotation in the outer region in this new theory. Thus, after giant impact on the surface of the differential proto-Earth, the materials in the debris disk originate mainly from the proto-Earth. Different from the resulting high angular momentum \citep{15} of the Earth-Moon system which should be removed to match the present-day value, the theory proposed in this work does not violate the angular momentum constraint when adopting a suitable low angular velocity of the inner region with a suitable radius.

The theory proposed above and that in \cite{15} try to obtain a circum-terrestrial debris disk made of materials mainly from the proto-Earth's mantle in order to account for the Moon's Earth-like isotopic composition. What happens if the materials in the mantle of Theia has an equation of state different from that of the materials in the mantle of proto-Earth? Since smaller bodies, like Vesta and Mars, typically have a greater proportion of iron in their mantles and smaller iron-rich cores compared with Earth \citep{15,26}, Theia may have an iron-rich mantle with a higher FeO content compared with Earth's mantle \citep{27}, meaning that the density of materials in Theia's mantle is larger than that in Earth's mantle under the same pressure. Thus, in the circum-terrestrial debris disk produced by the giant impact, the blobs of materials originating from Theia would sink to the lower layer after accreting onto the moon while the blobs of materials from proto-Earth float on the upper layer, especially on the surface. From this point of view, the Moon's Earth-like isotopic composition is a natural result of the canonical giant impact. Simulations with respect to this theory will be conducted in the near future. Combination of the theory proposed in this work with the origin of the lunar inclination proposed in \cite{28} may provide a new insight into the formation and evolution of the Moon.

\section{Discussions}
We propose a new theory for the origin of the Moon's Earth-like isotopic composition in this work. According to simulations of planetary accretion, planetesimals hit the surface of the proto-Earth and therefore increase the rotation rate of the surface of the proto-Earth, giving rise to the faster rotation rate of proto-Earth's surface than the inner region and transfer of angular momentum from surface to inner and resulting in a differential rotation of proto-Earth. With the condition of the differential rotating proto-Earth, the giant impact of a sub-Mars-sized body on a proto-Earth with a fast rotating outer region and a relative slow rotating inner region shows to be a viable way of producing a circum-terrestrial debris disk with material mainly from the proto-Earth, resulting that the Moon could have isotopic compositions identical to that of Earth.

It should be noted that, in this work, only one set of parameters is adopted in our numerical simulation. We also have already tested our mechanism by using several sets of parameters with much more particles and the results support the feasibility of this theory. Comprehensive parameter survey will be conducted to test the feasibility of this theory in the near future.

%\bibliographystyle{aasjournal}
%\bibliography{reference} % if your bibtex file is called example.bib

\begin{thebibliography}{}
\expandafter\ifx\csname natexlab\endcsname\relax\def\natexlab#1{#1}\fi
\providecommand{\url}[1]{\href{#1}{#1}}










\bibitem[Hartmann \& Davis(1975)]{1}Hartmann, W. K., \& Davis, D. R.\ 1975, Icarus, 24, 504
\bibitem[Cameron \& Ward(1976)]{2}Cameron, A. G. W., \& Ward, W. R.\ 1976, Lunar Planet. Sci. Conf., 7, 120
\bibitem[Canup \& Asphaug(2001)]{3}Canup, R. M. \& Asphaug, E.\ 2001, Nature, 412, 708
\bibitem[Canup(2004)]{4}Canup, R. M.\ 2004, Icarus, 168, 433
\bibitem[Canup(2008)]{5}Canup, R. M.\ 2008, Icarus, 196, 518
\bibitem[Ringwood(1986)]{6}Ringwood, A. E.\ 1986, Nature, 322, 323
\bibitem[Lugmair \& Shukolyukov(1998)]{7}Lugmair, G. W. \& Shukolyukov, A.\ 1998, Geochim. Cosmochim. Acta, 62, 2863
\bibitem[Wiechert et al.(2001)]{8}Wiechert, U., et al.\ 2001, Science, 294, 345
\bibitem[Touboul et al.(2007)]{9}Touboul, M., et al.\ 2007, Nature, 450, 1206
\bibitem[Zhang et al.(2012)]{10}Zhang, J., et al.\ 2012, Nature Geosci. 5, 251
\bibitem[Herwartz et al.(2014)]{11}Herwartz, D., et al.\ 2014, Science, 344, 1146
\bibitem[Young et al.(2016)]{12}Young, E. D. et al.\ 2016, Science, 351, 493
\bibitem[Meier(2012)]{13}Meier, M. M. M.\ 2012, Nat. Geosci. 5, 240
\bibitem[Mastrobuono-Battisti et al.(2015)]{14}Mastrobuono-Battisti, A., et al.\ 2015, Nature, 520, 212
\bibitem[{\'C}uk \& Stewart(2012)]{15}{\'C}uk, M., \& Stewart, S. T.\ 2012, Science, 338, 1047
\bibitem[Lock \& Stewart(2017)]{16}Lock, S. J., \& Stewart, S. T.\ 2017, JGRE, 122, 950
\bibitem[Lock et al.(2018)]{17}Lock, S. J., et al.\ 2018, JGRE, 123, 910
\bibitem[Canup(2012)]{18}Canup, R. M.\ 2012, Science, 338, 1052
\bibitem[Agnor et al.(1999)]{19}Agnor, C. B. et al.\ 1999, Icarus, 142, 219
\bibitem[Kokubo \& Ida(2007)]{20}Kokubo, E. \& Ida, S.\ 2007, Astrophys. J. 671, 2082
\bibitem[Kokubo \& Genda(2010)]{21}Kokubo, E. \& Genda, H.\ 2010, Astrophys. Lett. 714, L21
\bibitem[Stewart et al.(2020)]{22}Stewart, S. T., et al.\ 2020, AIP Conf. Proc. 2272, 080003
\bibitem[Ruiz-Bonilla et al.(2021)]{23}Ruiz-Bonilla, S., et al.\ 2021, Mon. Not. R. Astron. Soc. 2870, 2861
\bibitem[Kegerreis et al.(2019)]{24}Kegerreis, J. A., et al.\ 2019, Mon. Not. R. Astron. Soc. 487, 1536
\bibitem[Hosono et al.(2019)]{25}Hosono, N., et al.\ 2019, Nat Geosci. 12, 418
\bibitem[Fischer et al.(2024)]{26}Fischer, M., et al.\ 2024, Proc Nat Acad Sci, 121, 52
\bibitem[Yuan et al.(2023)]{27}Yuan, Q., et al.\ 2023. Nature, 623, 95




\bibitem[Liu(2025)]{28}Liu, W.\ 2025, arXiv:2502.18837


%\bibitem{14}A. A. Starobinsky, Sov. Astron. Lett., 11, 133 (1985)
%\bibitem{15}V. A. Rubakov, M. V. Sazhin, \& A. V. Veryaskin, Phys. Lett., 115B, 189 (1982)
%\bibitem{16}R. Fabbri \& M. D. Pollock, Phys. Lett. 125B, 445 (1983)












\end{thebibliography}

\end{document}